%% file: fea.nocomments.v022.tex
\documentclass{eptcs}
 \pdfoutput=1

\usepackage{ifthen}
\newboolean{commentsaon}         
\setboolean{commentsaon}{true}
  \setboolean{commentsaon}{false}  

\newboolean{commentson} 
\setboolean{commentson}{true}

\newcommand{\comment}[1]
{\ifthenelse{\boolean{commentson}\AND\boolean{commentsaon}}
   {{\par\noindent\mbox{}{\footnotesize\blue[ *** #1 ]\par}\noindent\par}}{}}

\newcommand{\commenta}[1]
{\ifthenelse{\boolean{commentsaon}}
   {{\par\noindent\mbox{}{\small\color[rgb]{0, .5, 0}[ *** #1 ]\par}\noindent\par}}{}}

\usepackage{myparagraph}   

\usepackage{xspace}
\usepackage{latexsym}
\usepackage{chemarrow}  

\usepackage{graphicx}
\usepackage{wrapfig}

\usepackage[yyyymmdd]{datetime}

\usepackage{amssymb}  %

\usepackage{color}
\newcommand\blue     {\color{blue}}

\newcommand{\grey}{\color[rgb]{.7, .7, .7}}

\newcommand{\mypalegrey}{\color[rgb]{.9, .9, .9}}
\newcommand\black {\color{black}}

\newcommand*{\seq}[2][n]  {{#2_{1}, \allowbreak \ldots, \allowbreak #2_{#1}}}

\newcommand*{\notmodels}{\mathrel{\,\not\!\models}}

\newcommand*{\A}{{\ensuremath{\cal A}}\xspace}
\newcommand*{\MP}{{\ensuremath{{\cal M}_P}}\xspace}

\newcommand*{\HB}{{\ensuremath{\cal H\!B}}\xspace}

\title{On Feasibility of Declarative Diagnosis}
\author{W{\l}odzimierz Drabent%
  \institute{Institute of Computer Science,  Polish Academy of Sciences}
  \email{drabent\,{\it at}\/\,ipipan\,{\it dot}\/\,waw\,{\it dot}\/\,pl%
    \rm
        {\ifthenelse{\boolean{commentson}\AND\boolean{commentsaon}}%
          {\blue\footnotesize\quad  [with private comments]}{}%
        }%
  }  
}  

\newcommand*\titlerunning{Declarative diagnosis}
\newcommand*\authorrunning{W. Drabent}

\begin{document}
\maketitle

\begin{abstract}
  The programming language Prolog makes declarative programming possible,
  at least to a substantial extent.  Programs may be written and reasoned
  about in terms of their declarative semantics.  All the advantages of
  declarative programming are however lost
 when it comes to program debugging.  This is because 
  the Prolog debugger is based solely on the operational
  semantics.  Declarative methods of diagnosis (i.e.\ locating errors in
  programs) exist, but are neglected.
  This paper discusses their possibly main weaknesses
 and shows how to overcome them.
  We argue that useful ways of declarative diagnosis of logic programs exist,
  and should be usable in actual programming.
\\[.5ex]
\end{abstract}

\section{Introduction}

The main concept of logic programming is that a program is a set
of formulae, and computation produces its logical consequences.
Such program can be understood declaratively -- not as a description of
any computation, but rather as a description of a problem to solve.
Its answers depend solely on its ``logic'', i.e.\ on the formulae of which it
consists.  So logic programming is a declarative programming paradigm.
The programmer can construct programs and reason about them at a higher level
of logic, abstracting from their computations.
Whole reasoning about program correctness (more precisely, correctness and
completeness) can be done at this level.
This is an important advantage of the paradigm.
One needs to resort to the operational semantics (the ``control'') only to
deal with termination and efficiency.  Modifying the control does not change
(the set of) the answers of the program.  What is changed is the way they are
computed; the logic is separated from the control
\cite{DBLP:journals/cacm/Kowalski79}.
The programming language Prolog was introduced as an implementation of logic
programming.  It is still the major logic programming language.
It makes declarative programming possible, at least to a substantial extent.  

On the other hand, a Prolog program can be looked at from a point of view of
its operational semantics.  A program is a precise description of
computations; one can program in Prolog imperatively.  This is often
necessary, for instance when a program has to interact with its environment.
Dealing with the operational semantics is often difficult, due to among
others the non-straightforward nature of backtracking, coroutining and
tabulation.

An important activity in programming is program debugging.
And when it comes to debugging of Prolog programs, 
all the declarativeness is lost. %
Declarative methods,
known as declarative diagnosis (DD) or algorithmic debugging,
exist but are neglected.
Basically, the only tool available for a programmer is the Prolog debugger,
which is based solely on the operational semantics.
So the tool is incompatible with declarative programming
(cf.\,\,e.g.\,\cite{lloyd87dd,DBLP:journals/jlp/DucasseN94,drabent.lopstr19}).
To use the debugger the programmer has to go into details of the operational
semantics, abandoning the declarative thinking.
This makes the debugging difficult, and encourages programmers to 
resign from 
the declarative view of programs.
The difficulties 
are even more serious when more sophisticated control is involved, like
coroutining or tabulation.

As the Prolog debugger is a rather powerful tool, 
one may expect that a ``declarative programmer'' can obtain from it the
information she needs.  An example of such information is which
premises have been used to derive a given answer, displayed at an {\tt exit}
port.
Obtaining such information is possible, but surprisingly tedious
\cite{drabent.lopstr19}.
\comment{
    The difficulties with the debugger 
    are even more serious with more sophisticated control, like coroutining
    or tabulation.
}

In the next section we briefly
introduce declarative diagnosis (DD),
present a unifying view of incorrectness and incompleteness diagnosis,
and suggest a kind of suitable DD tools.
Then (Section \ref{sec.intended.model}) we identify the intended
model problem, which has possibly been the main obstacle for acceptance of DD,
and show how to overcome it.
In Section \ref{sec.extracting}
 we discuss applying DD to Prolog with tabulation and delays.

\paragraph
{Preliminaries.}
The presentation
is rather informal, in most cases references
are given to a more precise treatment.
In some places the terminology of logic is used \cite{Apt-Prolog}, instead of
that of Prolog manuals.  So by
an {\em atom} we mean an atomic formula, not a Prolog constant.
We consider SLD-resolution dealing with {\em queries} (conjunctions of atoms),
instead of goals (negated queries).  
From a programmer's point of view, 
the selected atom in a query in an SLD-derivation is a {\em procedure call}.
By a ({\em computed}) {\em answer} of a program we mean
the result of applying a  (computed) answer substitution to the initial query.
The set of clauses (in a program) beginning with a predicate symbol $p$
will often be called {\em procedure} $p$.
The Herbrand base will be denoted by \HB, 
and the least Herbrand model of a program $P$ by \MP.

In imperative and functional programming, program correctness consists of
program termination and partial correctness.  The latter means that the
program produces required results provided it terminates.
In logic programming, due to its nondeterministic nature, partial  correctness
splits into
{\em correctness} and {\em completeness}
\cite{hogger.book,DBLP:journals/tplp/DrabentM05shorter,drabent.tocl16}.
  Correctness means that each answer
of the program is compatible with the specification.  Completeness means that
all the answers required by the specification are produced by the program.
Obviously, termination is a property of the operational semantics,
in contrast to correctness and completeness.
Debugging consists of program {\em diagnosis}, i.e.\ locating errors in
programs, and program correction.
In this paper we are interested in diagnosing incorrectness and
incompleteness for logic / Prolog programs.

\section{Declarative diagnosis}
\label{sec.DD}

The idea of declarative diagnosis was introduced by 
Shapiro in his seminal thesis \cite{shapiro.book}, under the name 
algorithmic debugging. 
Shapiro gave algorithms for diagnosing incorrectness and incompleteness
(and also non-termination).  His ideas were followed and developed
by further work, dealing mainly with logic programming, but  also with other
paradigms (see e.g.\ 
\cite{%
Pereira86-short,
DNM89,
{DBLP:journals/ngc/Naish92},
  csur/CaballeroRS17}).

\paragraph{Specifications, symptoms, errors.}
Diagnosis starts with a {\em symptom} of program incorrectness or
incompleteness.
The purpose is to locate in a program an {\em error},
i.e.\,a fragment responsible for the symptom.
In incorrectness diagnosis a symptom is an incorrect answer of the program.
In diagnosing incompleteness it is, generally speaking, a missing answer.
Here by an {\em incompleteness symptom} we mean an atomic query for which 
the program terminates, 
but it does not produce some required answers.

Diagnosis must be based on a specification.
In DD the specification is the {\em intended model} of the program.
This is
 the least Herbrand model of 
the target program.
A DD algorithm is informed about the intended model by an
{\em oracle}, which answers queries about the model.  Usually the oracle is
the user.   %
\enlargethispage{1ex}
\smallskip
\[
{%
\begin{array}{c c c c}
  \begin{tabular}{r}
\raisebox{1ex}[16pt] %
         {%
    \begin{tabular}{c}
      program \\[-.7ex] +\\[-1ex] symptom
    \end{tabular}
    \quad $\rarrowfill{3em}$%
         }  %
    \\[5ex]
    located error     \ \ \,$\larrowfill{3em}$\,  %
  \end{tabular}
&
  \fbox{
    \rule[-4.5ex]{0pt}{10ex}%
   \ DD algorithm \
  }
&
  \begin{array}{cc}
    \autorightarrow{queries}{}
    \\[-1ex]
    \mbox{\small\footnotesize%
      \begin{tabular}{c}
        about the \\ intended model
      \end{tabular}%
    }
    \\
    \autoleftarrow{\hspace*{3em}}{answers}
  \end{array}
&
  \fbox{
    \rule[-4.5ex]{0pt}{10ex}%
    \quad\ 
    \begin{tabular}{c}
    \mbox{\large user}      
    \\
    \small\footnotesize (oracle)
    \end{tabular}
    \quad\ 
  }
  \end{array}
} %
\]
In incorrectness diagnosis, an {\em incorrectness error}
is an incorrect clause.  
In such clause the body atoms are true, and the head is not true in the
intended specification.  In logical terms $M\notmodels C$, where
$M$ is the intended model and $C$ the clause.

By an {\em incompleteness error} we will mean an atomic query $A$ 
 whose answer
(required by the specification)
cannot be produced by any clause of the program
(out of atomic queries required by the specification).
Logically, we have an instance $A\theta\in M$ such that 
the program does not contain a clause with an instance 
    $A\theta\gets\vec B$ such that $M\models\vec B$, where $M$ is the intended model.
So an
incompleteness error points at a procedure responsible for the error
(and presents an atom the procedure should additionally produce).

It can be shown that a more precise notion of an error is impossible.  For
instance, an incorrect clause can be corrected in various ways, leaving
various its fragments unchanged.  So we cannot consistently name such
a fragments as correct, or incorrect.

\paragraph{DD algorithms, a unifying view.}
For incompleteness, we prefer Pereira style diagnosing
\cite{DBLP:journals/ngc/Naish92}, as it is more convenient than
that of  \cite{shapiro.book}.
In such context, both incorrectness and incompleteness
  diagnosis can be treated as instances of a single algorithm.
(See \cite{DBLP:journals/jflp/Naish97,csur/CaballeroRS17} for a similar
  observation.)

Let us first describe the oracle queries.  A query in incorrectness
diagnosis is an atom $A$ (which was obtained as an answer in the computation).
In incompleteness diagnosis it is an atom $A$ (a procedure call from the
computation) together with the computed answers $\seq{A\theta}$ ($n\geq0$)
obtained for
$A$.
The oracle has to answer whether the query is a symptom
(of correctness or completeness, respectively).

The algorithm 
inspects an erroneous computation of the program, and builds
builds a {\em DD search tree} of oracle queries.  The root is
the symptom we begin with.
The tree contains a {\em target},
i.e.\ a node which is a symptom and its children are not.  
The algorithm employs the oracle to search the tree for a target. 
Now we %
give further details of the tree and
 explain how a target determines an error. 

In incorrectness diagnosis the tree is a proof tree (the root is a wrong
answer, 
each node together with
its children is an instance of a program clause).  Thus a target with its
children is an incorrectness error.

For incompleteness diagnosis, each node \A
(which is an atom $A$ with its computed answers)
has children that are the top-level queries (together with the answers) used
in the computation for $A$.  
(By top-level we mean one that is an instance of a body atom
of a clause used to resolve  $A$.)
If \A is a target and its children are not,
then \A contains an incompleteness error
(which is the procedure call from \A).
Informally, as we found no incompleteness related to the body atoms of the
applicable clauses, the clauses are responsible (for the initial symptom).

It is important that with a DD algorithm the user can locate the error 
without looking at the program.  Also, all the diagnosis is performed solely in
terms of the declarative semantics.  So we do not need to understand the
operational semantics.

\paragraph{Non-algorithmic declarative diagnosis.}
Practice shows some inconveniences of using DD algorithms,
at least in their basic form (cf.\,e.g.\,\cite{DNM89}).
An inconvenient and counterproductive feature is that 
the order of queries is determined by the algorithm.
The effort (and time) to answer various queries may differ substantially,%
\footnote{
  So it is not realistic to judge the efficiency of DD algorithms by means 
  of query complexity \cite{shapiro.book,csur/CaballeroRS17},
  which deals solely with the number of queries.
}
so it is desirable to first deal with queries considered by the user
to be easier.  As a result, some difficult queries many eventually not be asked.
Also, 
it should be possible to modify one's previous answer, as errors in answering
are possible.  Such modifications are also useful in making and withdrawing
temporary assumptions (``what if this were correct'').

Such inconveniences disappear when the search is performed by the user.
It is sufficient that a DD tool extracts from the computation 
the DD search tree, and gives the user convenient means to explore the tree.
The author's experience
with such prototype tools
shows that they are more convenient to use than (prototype implementations of)
DD algorithms. 
The main burden -- with understanding and answering queries -- is the
same (only the answers are not typed in).   Additional 
human effort, due to performing the search, is minor.
Being relieved of the restraints of DD algorithms is a substantial
convenience.
The user may look at various queries and choose which to answer first.
Some answers may be postponed, the user may inspect the tree freely.
An already visited part of the tree may be re-inspected, to correct former
decisions (if erroneous, or assumed temporarily).

What we propose is, in a sense, a declarative counterpart of the Prolog
debugger.
The debugger gives the user access to the operational semantics of a program;
we show how to give access to the declarative semantics.

\enlargethispage{.9ex}

\section{The intended model problem}
\label{sec.intended.model}

In this section, based on  \cite[Section 7]{drabent.tocl16},
 we discuss what possibly is the main reason for lack of
acceptance of DD in practice.
Namely a fundamental notion of DD is that of the intended model.
This means the the programmer has to know exactly the intended least Herbrand
(2-valued)
model of her program;  in other words, to know exactly the relation to be
defined by each predicate.   This is usually not realistic.

As an example \cite{drabent.tocl16}
 consider insertion sort and predicate insert/3, dealing with
inserting an element into a sorted list (to obtain a sorted list as a result).
The user does not (and should not) know how the predicate should behave on
unsorted lists.  Hence she is unable to answer an oracle query like
``Is atom $B={\it insert}(2,[3,1],[2,3,1])$ correct''?
Let us call this phenomenon the {\em intended model problem}.

Both treating $B$ as correct, or incorrect may be reasonable, and result 
in different debugged programs.  For instance treating all such atoms as
incorrect leads to a program in which insert/3 checks that its second
argument is a list.  Such program is inefficient.
So choosing the intended model may  then force certain 
design decisions, possibly undesirable ones.
Often while developing a program it is impossible to know in advance its exact
semantics (i.e.\ its intended model).
Apparently the semantics of insert/3 (in a standard insertion sort program
\cite[Program 3.21]{Sterling-Shapiro-shorter})
is a result of making the program simple and efficient, and has not been
decided in advance.
See \cite{drabent.tplp18} for a larger example.

In the author's opinion the intended model problem
makes DD inapplicable in many, if not in most of, practical cases.

\paragraph{Approximate specifications.}
Such examples show a common feature of specifications in logic programming
(both formal, and informal ones which
programmers have in their minds): often such
 specifications 
have a 3-valued flavour.   
Some atoms are clearly true, some false, and the
remaining ones are%
{\setlength{\parfillskip}{0pt}%
\setlength{\parskip}{0pt}%
\setlength{\emergencystretch}{.5\textwidth}%
\par}
\vspace{-.5ex}
\input{diagram-approximate}\noindent
irrelevant. %
They may be accepted as answers, %
but should not be required to be answers of the program.

\newcommand{\myspc}{\hspace{-3pt}}  

So instead of specifying a single intended model, we should provide two
specifications,  $S^0$ for completeness and $S$ for correctness
(where $S^0\!\subseteq S$).
The irrelevant atoms, like $B$ %
in our example are in $S\setminus S^0$.
Let us call such pair
$S^0\myspc,S$ of specifications an {\em approximate specification}. 

A program $P$ is correct w.r.t.\ an approximate specification $S^0\myspc,S$ 
if $S\models A$ for each 
answer $A$ of $P$ (i.e.\ $P\models A$ implies $S\models A$);
$P$ is complete w.r.t.\,it if $S^0\!\models A$ implies that $A$ is an answer of $P$
($S^0\!\models A$ implies $P\models A$).

A reasonable approximate specification for insert/3 of insertion sort is:
\[
\hspace{-3em}
\begin{array}{ll}
\mbox{for completeness:} &
     S^0_{\it insert} = \left\{\,
          \begin{array}{@{}r@{}}
              {\it insert}( n, l_1, l_2) \in\HB
            \end{array}
              \ \left| \ 
            \mbox
          {\begin{tabular}{@{}l@{}}
              $l_1, l_2$ are sorted lists of integers, \\
              ${\it elms}(l_2) = \{n\}\cup {\it elms}(l_1)$
            \end{tabular}
          }
        \right.\right\},\hspace*{-1em}
\\ [2ex]
&
\hspace{14em}
\makebox[0pt][l]
    {\footnotesize
      where ${\it elms}(l)$ -- the multiset of elements of $l$,%
    }
\\ [.5ex]
\mbox{for correctness:} &
    S_{\it insert} = \left\{\,
          \begin{array}{@{}r@{}}
              {\it insert}( n, l_1, l_2) \in\HB %
            \end{array}
              \ \left| \ 
        \mbox
          {\begin{tabular}{@{}l@{}}
              if $n$ is an integer and
              \\ $l_1$ is a sorted list of integers, \\
              then ${\it insert}( n, l_1, l_2)\in S^0_{\it insert}$
            \end{tabular}
          }
        \right.\right\}.
\end{array}
\hspace{-3em}
\]
Note that neither $S^0_{\it insert}$, nor $S_{\it insert}$ corresponds 
to the semantics of the actual corrected insertion sort program
(i.e. neither of them is 
the set of atoms of the form ${\it insert}(\vec t)$ from its
 least Herbrand model).

Obviously, diagnosis of incorrectness and that of incompleteness should be
done employing the respective specification.
With such approach to specifications,
the standard DD methods (based on intended, 2-valued interpretations) are
sufficient.

The intended model problem has been noticed, among others, by 
\cite{Pereira86-short,DNM89,DBLP:conf/acsc/Naish00,DBLP:conf/aadebug/Ferrand93}.
Pereira \cite{Pereira86-short} introduces inadmissible atoms, which should
not appear in the computation.  In this way the debugging is no more
declarative as it refers also to the intended operational semantics of the
program. 

Naish \cite{DBLP:conf/acsc/Naish00} introduces a rather sophisticated
3-valued approach to DD, and two kinds of errors (e-bug and i-bug).
The atoms from $S_{\it insert}\setminus S^0_{\it insert}$
above
have in the 3-valued intended model the third logical value {\bf i}.
(Those from $S^0_{\it insert}$ are {\bf t}, and those from 
 $\HB\setminus S_{\it insert}$ are {\bf f}.)
It seems strange, that both incorrectness and incompleteness diagnosis treat
in the same way the {\bf i} atoms;  the discussion above shows that this
should not be the case.

Any such complications are not necessary 
when we have separate specifications for correctness and completeness.
Introducing such specifications solves the intended model problem.
In the author's opinion this is a main step to make DD practical.

\paragraph{Negation and partial specifications.}
When we deal with logic programs with negation, a specification should also
describe the negative results of programs; roughly, which ground atoms fail.
In other words, which negated literals are program consequences under
the chosen semantics.

\commenta{Kunen semantics, well-founded semantics \ \ \ Mention ASP}
\comment{29/4: ...
   Prolog augmented with negation as failure implements the 3-valued
   completion semantics by Kunen \cite{Kunen87}, provided
   that floundering is avoided \cite{drabent96}.
    Floundering means here selecting a non-ground
   negative literal, which neither finitely fails, nor produces a most general
   answer,   When tabulation is introduced, an infinite SLD-tree with no answers
   may be inspected in a finite time.  Hence Prolog with tabulation approximates
   the well-founded semantics.
}

As in the discussion above, a specification often has to distinguish between
those atoms that have to fail (specifying completeness)
from those that may fail (specifying correctness).
To specify both positive and negative program answers,
it is natural to use an approximate specification $S^0\myspc,S$ as introduced above,
and interpret it as follows \cite{DBLP:journals/tplp/DrabentM05shorter}.
The atoms that are not allowed to succeed
have to
fail.  The atoms that are required to succeed cannot fail.
Thus the set of atoms that have to fail is $\HB\setminus S$
(the ``erroneous'' in the diagram).
Similarly, the set of atoms that may fail is $\HB\setminus S^0$
(the ``irrelevant'' and ``erroneous'' in the diagram).
In a correct program, if $\neg A$ is its answer then $S^0\!\models\neg A$.
In a complete program, if $S\models\neg A$. then $\neg A$ is an answer.
So the description of correctness for positive answers describes completeness
of negative ones, and vice versa.

We do not discuss here DD of programs with negation
\cite{lloyd87dd,DNM89,snt.meta90}.
We only mention that finding $\neg A$ as a target in incorrectness diagnosis
leads to incompleteness diagnosis for $A$ (and vice versa).

\section{Extracting the DD search tree.}
\label{sec.extracting}
This section considers implementing declarative diagnosis for Prolog with
tabulation 
and coroutining.  It turns out that only incompleteness diagnosis for
coroutining poses problems.  We discus two ways of dealing with them.

DD requires extracting the necessary information (i.e.\ the DD search tree)
from a computation of the program.  For basic Prolog (without tabulation and
coroutining) we know how to do this, see e.g.\ 
\cite{shapiro.book, DNM89,DBLP:journals/ngc/Naish92}; we skip further details.
For incorrectness diagnosis this means extracting the proof tree.
(Most of) the methods to do this for basic Prolog seem easy to generalize 
for coroutining and tabulation.
Such generalization is far from obvious for incompleteness diagnosis.

For incompleteness diagnosis (in Pereira-style) one needs to collect 
from the computation the procedure calls, and for each of them the obtained
computed answers.  
For basic Prolog,
a simplest way to do this
seems
top-level meta-interpretation for the considered procedure call
(cf.\ \cite{DNM89}, where however the DD search tree is not made explicit).
So for a call $p(\vec t)$, all the clauses $p(\vec s)\gets\seq B$
are meta-interpreted, however each encountered procedure call $B_i\theta_j$ is 
executed by the Prolog system.
All such calls (together with
their answers) are children in the DD search tree of the node $p(\vec t)$ (with
its answers).

Other ways of obtaining the search tree for incompleteness DD may be
suitable.  In particular, obtaining the whole tree from a single computation  
may be preferable (as this mimics the actual side effects of a
not purely declarative program).

Transferring such ideas to Prolog with tabulation should not create problems.
The task is however difficult for programs with coroutining.
This is because
the answers for a given procedure call may actually not be computed
(during the computation for the considered initial query).
The query instance returned by Prolog as an answer (e.g.\ that appearing
at the Exit port of Prolog debugger)
will be called here a {\em pseudo-answer}.
See \cite{drabent2023implementing.arxiv} for a formal description of
coroutining.

Let us explain.  During a computation for a sub-query $A$
it may happen that
(i) unblocking of a formerly delayed sub-query results in a
pseudo-answer being an instance of an actual computed answer, or
(ii)
delaying of a sub-query results in a pseudo-answer
being more general than an actual one (or an actual answer may
not exist).
When %
(i) and (ii) coincide, nothing can be concluded.

For sound incompleteness diagnosis, a DD query should contain actual computed
answers for a given sub-query $A$, but what we can obtain from the
computation are pseudo-answers. 
This difficulty was discussed in  \cite{DBLP:journals/ngc/Naish92}, without a 
general solution.
Executing $A$ without coroutining is of no help in general, as it may result in
non-termination. 
It is useful to do this under a suitable time limit;
if $A$ terminates, the problem is solved.  
For a general case consider the following.
From a computation trace one can find if cases (i) or (ii) have
been involved in producing a given pseudo answer for $A$.
If not, the pseudo-answer is an answer.  
If (i) has not occurred then
the pseudo-answer is more general than the actual answer (or the latter does
not exist).  
If (ii) has not occurred, the pseudo-answer is an instance of an
actual one.
A pseudo-answer together with such information can be included in a DD query,
instead of the unknown actual answer; the information makes it possible in
some cases to answer the query (formally, to state whether $A$ with the actual
answers is an incompleteness symptom).%
\footnote
{\label{footnote.i.ii}
If case (i) has not occurred for any pseudo-answer for $A$, and
some required answer is not an instance of any displayed (pseudo-)answer,
then $A$ (with its possibly unknown answers) is surely a symptom.
If case (ii) has not occurred for any pseudo-answer for $A$,
and each answer required by the specification is an instance of some
displayed one,
then $A$ (with answers) is surely not a symptom.
} 

This approach may fail to find that, despite of the involved coroutining,
 a pseudo-answer is an answer.  We propose another method of checking 
this, and to find the actual answer in some cases.

From the computation producing a pseudo-answer $A\theta$ for $A$
(w.r.t.\ a program $P$)
a proof tree may be collected as in incorrectness diagnosis.  
Due to delays, some subtrees of an actual proof tree may be missing,
so the obtained %
tree will be called a {\em pseudo-proof tree}.
Such tree $T$, with root $A\theta$,
is inspected to find if it is (a) a proof tree or (b) not.
For (a), $A\theta$ is an answer of $P$, 
%
but it may not be a computed answer for $A$ (but its instance, case (i)).
Out of the tree we can obtain the clauses of $P$ used to compute $A\theta$.
Applying the clauses, we can obtain a computed answer $A\sigma$ for $A$, so that
$T$ is an instance of the proof tree $T'$ for $A\sigma$
(and $A\sigma$ is more general than $A\theta$).
Now $A\sigma$ can be used in a DD query,

The same procedure can be applied to $T$ in case (b), 
it results in a pseudo-proof
tree $T'$ with root $A\sigma$, more general than $A\theta$.
The tree corresponds to executing $A$ alone (in a context without any pending
delayed sub-queries).
Formally, $T'$ corresponds to a prefix $D$ of an SLD-derivation for $A$;
the last query $Q$ of $D$ consists of the delayed sub-queries.
So any computed answer for $A$ obtained by completing the derivation
is an instance of $A\sigma$. 
Thus $A\sigma$ can be used instead of the unknown answer in a DD query,
and be treated like a pseudo-answer for which (i) has not
occurred (cf.\ footnote \ref{footnote.i.ii}).

\commenta{Not really true:
(The atoms of $Q$ are those leaves of $T'$ that are not instances of facts of
the program). 
}

As previously, it makes sense to execute $Q$ under a time limit.  
If it terminates, 
then all the derivations with $D$ as a prefix have been found.
They provide computed answers for $A$, to be used in a DD query
(instead of the pseudo-answer $A\theta$).

To summarize, generalizing DD to take care of tabulation and coroutining 
seems rather obvious, except for diagnosing incompleteness for coroutining.
In this case some DD queries cannot be extracted from the computation of the
program. 
We discussed how to deal with this problem.
For most of cases we showed a way
to obtain a sound answer for such an undetermined DD query.
There is one exception (in case
(b) above, when the delayed $Q$ does not terminate under a given time limit,
one cannot determine that the DD query is not a symptom).
We do not pursue here
another possible approach,
based on detailed monitoring of variable
bindings during the computation.

\section{Conclusion}

\paragraph
{Comments.}
Due to lack of space, we do not discuss various issues related to pragmatics
of DD (declarative diagnosis), and to Prolog features apart basic ones.
The importance of a user friendly interface is obvious (it should be possible
to view big terms effectively, to conveniently inspect DD search trees, etc).
Experience from actual debugging is needed to evaluate diagnosis approaches,
and to better understand which features a useful diagnosis tool should have.
The author is interested in buggy programs that could be used in such
experiments, preferably with challenging or otherwise interesting bugs.

DD diagnosis turns out to be applicable also to programs with
non-declarative fragments.
Of course this concerns only issues related to the program answers (and not
e.g.\,the sequences of input/output actions).
For instance, prototype ``non-algorithmic'' DD tools
(cf.\,Section\,\ref{sec.DD}) for incorrectness and
incompleteness have been used to debug themselves.
Locating errors with these tools seems substantially simpler than with the
Prolog debugger.

Debugging for ASP 
is a separate issue, not dealt with here (as the role of answers in
ASP is different from that in standard logic programming).
An interesting challenge is to generalize DD so that more symptoms (of 
possibly a single error)
can be used to locate the error more efficiently.

\paragraph{Summary.}
This note deals with DD of logic programs.
We treat incorrectness diagnosing and incompleteness diagnosing
(Pereira style) as instances of the same algorithm.
We point out the intended model problem as possibly the main reason
for lack of acceptance of DD in practice.
We also consider DD
for programs with tabulation and
coroutining, and propose how to cope with the difficulties that arise.
We advocate that a tool which allows the user to inspect
the DD search tree is more suitable to perform diagnosis than an
implementation of a whole DD algorithm.

\bibliographystyle{eptcsalpha}

\bibliography{bibshorter,bibmagic,bibpearl,dd}

\end{document}

%% file: diagram-approximate.tex

\begin{wrapfigure}{r}{0.55\textwidth}
\hfill
\begin{minipage}{.4\textwidth}
$  
%
\newlength{\unit}\newlength{\vunit}  
\setlength{\unit}{.7cm}
   \setlength{\vunit}{\unit}
\begin{array}{l}
    \grey
     \overbrace{\rule{4.5\unit}{0pt}}^
         {\raisebox{.5ex}{\makebox[0pt][c]
             {\footnotesize\black \MP}}
         }
     \\
\hspace{1pt}%
\overbrace{\rule{2.9\unit}{0pt}}^
    {\raisebox{.5ex}{\makebox[0pt][c]
        {\footnotesize specification for completeness}}
    }
\\ [-1ex]
\framebox[3\unit]{\parbox[t][2\vunit][c]{3\unit}{ \hfil required\hfil }}
\hspace{-.5pt}
\rlap{\mypalegrey\vrule height.15\vunit depth2.15\vunit width2.98\unit}%
\rlap{\hspace{1.55\unit}\raisebox{-2.18\vunit}[0pt][0pt]{%
                     \makebox[0pt][c]{%
                  \includegraphics[trim=4.8cm 24cm 16.4cm 2.1cm,clip,scale=.96]
                   {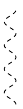}%
                   }}
}
\framebox[3\unit]{\parbox[t][2\vunit][c]{3\unit}
    { \hfil \raisebox{.5ex}{irrelevant}\hfil }%
}  
\hspace{-.5pt}
\rlap{%
\color[rgb]{1, 0.5, 0.5}
      {\vrule height.15\vunit depth2.15\vunit width2.98\unit}%
}%
\framebox[3\unit]{\parbox[t][2\vunit][c]{3\unit}{ \hfil erroneous\hfil }}
 \makebox[0pt][l]{%
     \raisebox{-1.15\vunit}
              {\ $\left.\rule{0pt}{1.3\vunit}\right\}\HB$}%
     }
\\ [-1ex]
\hspace{1pt}%
\underbrace{\rule{5.9\unit}{0pt}}_
     {\raisebox{-1ex}{\makebox[0pt][c]
         {\footnotesize specification for correctness}}
     }
\end{array}
$    
\end{minipage}%
\hspace*{3.5em}

\end{wrapfigure}%